\documentclass[aps,prl,preprint,tightenlines,superscriptaddress,showpacs,byrevtex]{revtex4}
\usepackage{graphicx} % Include figure files
\usepackage{dcolumn}  % Align table columns on decimal point

\graphicspath{{ps}}

\renewcommand{\arraystretch}{1.1}

\begin{document}

%\vspace*{-3\baselineskip}
%\resizebox{!}{3cm}{\includegraphics{belle.eps}}

\preprint{\vbox{ \hbox{   }
                 \hbox{BELLE-CONF-0735}
               % \hbox{ICHEP2006-xx}
               % \hbox{hep-ex nnnn, if available}
}}

\title{ \quad\\[0.5cm] Measurement of the inclusive 
\boldmath{$B_s^0 \to X^+ \ell^- \nu$} semileptonic decay branching fraction}

\affiliation{Budker Institute of Nuclear Physics, Novosibirsk}
\affiliation{Chiba University, Chiba}
\affiliation{University of Cincinnati, Cincinnati, Ohio 45221}
\affiliation{Department of Physics, Fu Jen Catholic University, Taipei}
\affiliation{Justus-Liebig-Universit\"at Gie\ss{}en, Gie\ss{}en}
\affiliation{The Graduate University for Advanced Studies, Hayama}
\affiliation{Gyeongsang National University, Chinju}
\affiliation{Hanyang University, Seoul}
\affiliation{University of Hawaii, Honolulu, Hawaii 96822}
\affiliation{High Energy Accelerator Research Organization (KEK), Tsukuba}
\affiliation{Hiroshima Institute of Technology, Hiroshima}
\affiliation{University of Illinois at Urbana-Champaign, Urbana, Illinois 61801}
\affiliation{Institute of High Energy Physics, Chinese Academy of Sciences, Beijing}
\affiliation{Institute of High Energy Physics, Vienna}
\affiliation{Institute of High Energy Physics, Protvino}
\affiliation{Institute for Theoretical and Experimental Physics, Moscow}
\affiliation{J. Stefan Institute, Ljubljana}
\affiliation{Kanagawa University, Yokohama}
\affiliation{Korea University, Seoul}
\affiliation{Kyoto University, Kyoto}
\affiliation{Kyungpook National University, Taegu}
\affiliation{\'Ecole Polytechnique F\'ed\'erale de Lausanne (EPFL), Lausanne}
\affiliation{University of Ljubljana, Ljubljana}
\affiliation{University of Maribor, Maribor}
\affiliation{University of Melbourne, School of Physics, Victoria 3010}
\affiliation{Nagoya University, Nagoya}
\affiliation{Nara Women's University, Nara}
\affiliation{National Central University, Chung-li}
\affiliation{National United University, Miao Li}
\affiliation{Department of Physics, National Taiwan University, Taipei}
\affiliation{H. Niewodniczanski Institute of Nuclear Physics, Krakow}
\affiliation{Nippon Dental University, Niigata}
\affiliation{Niigata University, Niigata}
\affiliation{University of Nova Gorica, Nova Gorica}
\affiliation{Osaka City University, Osaka}
\affiliation{Osaka University, Osaka}
\affiliation{Panjab University, Chandigarh}
\affiliation{Peking University, Beijing}
\affiliation{University of Pittsburgh, Pittsburgh, Pennsylvania 15260}
\affiliation{Princeton University, Princeton, New Jersey 08544}
\affiliation{RIKEN BNL Research Center, Upton, New York 11973}
\affiliation{Saga University, Saga}
\affiliation{University of Science and Technology of China, Hefei}
\affiliation{Seoul National University, Seoul}
\affiliation{Shinshu University, Nagano}
\affiliation{Sungkyunkwan University, Suwon}
\affiliation{University of Sydney, Sydney, New South Wales}
\affiliation{Tata Institute of Fundamental Research, Mumbai}
\affiliation{Toho University, Funabashi}
\affiliation{Tohoku Gakuin University, Tagajo}
\affiliation{Tohoku University, Sendai}
\affiliation{Department of Physics, University of Tokyo, Tokyo}
\affiliation{Tokyo Institute of Technology, Tokyo}
\affiliation{Tokyo Metropolitan University, Tokyo}
\affiliation{Tokyo University of Agriculture and Technology, Tokyo}
\affiliation{Toyama National College of Maritime Technology, Toyama}
\affiliation{Virginia Polytechnic Institute and State University, Blacksburg, Virginia 24061}
\affiliation{Yonsei University, Seoul}
  \author{K.~Abe}\affiliation{High Energy Accelerator Research Organization (KEK), Tsukuba} % KEK
  \author{I.~Adachi}\affiliation{High Energy Accelerator Research Organization (KEK), Tsukuba} % KEK
  \author{H.~Aihara}\affiliation{Department of Physics, University of Tokyo, Tokyo} % Tokyo
  \author{K.~Arinstein}\affiliation{Budker Institute of Nuclear Physics, Novosibirsk} % BINP
  \author{T.~Aso}\affiliation{Toyama National College of Maritime Technology, Toyama} % Toyama
  \author{V.~Aulchenko}\affiliation{Budker Institute of Nuclear Physics, Novosibirsk} % BINP
  \author{T.~Aushev}\affiliation{\'Ecole Polytechnique F\'ed\'erale de Lausanne (EPFL), Lausanne}\affiliation{Institute for Theoretical and Experimental Physics, Moscow} % ITEP
  \author{T.~Aziz}\affiliation{Tata Institute of Fundamental Research, Mumbai} % Tata
  \author{S.~Bahinipati}\affiliation{University of Cincinnati, Cincinnati, Ohio 45221} % Cincinnati
  \author{A.~M.~Bakich}\affiliation{University of Sydney, Sydney, New South Wales} % Sydney
% ...
% Lausanne

\collaboration{Belle Collaboration}
\noaffiliation

\begin{abstract}
Inclusive semileptonic $B_s^0 \to X^+ \ell^- \nu$ decays are studied
for the first time using a 23.6\,fb$^{-1}$ data sample collected
on the $\Upsilon$(5S) resonance with the Belle
detector at the KEKB asymmetric energy $e^+ e^-$ collider.
These decays are identified by the means of a lepton
accompanied by a same-sign $D_s^+$ meson originating 
from the other $B_s^0$ in the event.
% to suppress backgrounds due to non-strange $B$ decays.
The semileptonic branching fractions are measured in the
electron and muon channels to be
\mbox{${\cal B}(B_s^0 \rightarrow X^+ e^- \nu)\, =$}
\mbox{$(10.9 \pm 1.0 \pm 0.9)\%$} and
\mbox{${\cal B}(B_s^0 \rightarrow X^+ \mu^- \nu)\, =$}
\mbox{$(9.2 \pm 1.0 \pm 0.8)\%$}, respectively.
Assuming an equal electron and muon production rate
in $B_s^0$ decays, a combined fit yields
an average leptonic branching fraction
\mbox{${\cal B}(B_s^0 \rightarrow X^+ \ell^- \nu)\, =$}
\mbox{$(10.2 \pm 0.8 \pm 0.9)\%$}.
\end{abstract}


\maketitle

%%%% >>>> keep the final version single-spaced
\tighten

{\renewcommand{\thefootnote}{\fnsymbol{footnote}}}
\setcounter{footnote}{0}

Although $B_s^0$ mesons have been studied by LEP and Tevatron experiments
for a long time, only a few $B_s^0$ decay branching 
fractions have been measured \cite{pdg}.
In addition to these very high energy experiments,
an alternative source of $B_s^0$ mesons has been explored
recently,
where $B_s^0$ mesons are produced by $e^+ e^-$ colliders running 
at the $\Upsilon$(5S) energy.
First evidence for inclusive and exclusive $B_s^0$ 
decays at the $\Upsilon$(5S) was found by the CLEO 
collaboration \cite{cleoi,cleoe}, which collected a 
data sample of 0.42~fb$^{-1}$, and by the Belle 
collaboration \cite{beli,bele}, which used a data sample 
of 1.86\,fb$^{-1}$. However, the statistical significance of these 
first measurements was very limited.
In 2006 the Belle collaboration collected 
an additional 21.7~fb$^{-1}$ of data at the $\Upsilon$(5S),
allowing one to study several $B_s^0$ decay modes with 
improved statistical accuracy.

One of the most interesting $B_s^0$ decay modes, for which the
branching fraction
can be measured for the first time using the new $\Upsilon$(5S) data,
is the total inclusive semileptonic decay $B_s^0 \rightarrow X^+ \ell^- \nu$.
Charge-conjugate modes are implicitly implied everywhere in this paper.
The $e^+e^-$ collisions at the $\Upsilon$(5S) resonance
are well-suited for the ${\cal B}(B_s^0 \rightarrow X^+ \ell^- \nu)$ 
measurement, while it is probably unfeasible at very high energy colliders.
% This branching fraction is not yet measured and is 
% an important parameter in $B$ meson physics.

    Historically, the total inclusive semileptonic 
$B^{(0/+)} \rightarrow X \ell \nu$ decay branching fraction was measured many 
years ago using data obtained at the $\Upsilon$(4S) (see PDG for the full 
list of the measurements \cite{pdg}). The measured values covered the range 
of 9.7 to 11.0$\%$ (the most recent PDG average value is (10.78 $\pm$ 0.18)\%) 
and were in a poor agreement with the theoretically predicted 
branching fraction of about 12\% \cite{bigi}. Later, explanations 
for this difference were proposed in several theoretical 
papers \cite{falk,bagan,volosh,neubert}; however this discrepancy 
is not yet completely understood.

Measurements of the total semileptonic $B^0$, $B^+$ and $B_s^0$ branching 
fractions, together with corresponding well-measured lifetimes,
determine the semileptonic widths.
These widths for the $B^0$, $B^+$ and $B_s^0$ mesons are expected to be
equal, neglecting small corrections due to electromagnetic and
light quark mass difference effects.
If any significant difference between the $B^0$, $B^+$ and $B_s^0$
semileptonic widths were observed, it would indicate
an unknown source of lepton production in $B$ decays.

The correlated production of a $D_s^+$ meson and a same-sign
lepton at the $\Upsilon$(5S) resonance is used in this analysis 
to measure ${\cal B}(B_s^0 \rightarrow X^+ \ell^- \nu)$.
%The analysis is based on a 23.6 fb$^{-1}$ data sample recorded at 
%the $\Upsilon$(5S) resonance with the Belle detector at the 
%KEKB $e^+e^-$ energy-asymmetric collider.
The method exploits the fact that, at the $\Upsilon$(5S),
the dominant production of a $D_s^+$ meson and a same-sign fast lepton 
comes from the $B_s^{(*)} \bar{B}_s^{(*)}$ state.
Neither the $c\bar{c}$ continuum nor $B^{(*)}\bar{B}^{(*)}$ states
(except for a small contribution due to $\sim 19\%$ $B^0$ mixing effect) 
can result in a same-sign $c$-quark (i.e.,\ $D_s^+$ meson) and 
primary lepton final state. 
Simultaneous production of a same-sign $D_s^+$ meson and $\ell^+$ 
from one $B_s^0$ decay is totally negligible.

\section{Belle detector and data samples}

The data were collected with the Belle detector at KEKB \cite{kekb},
an asymmetric energy double storage ring collider with $\sim$8.2 GeV 
electrons and $\sim$3.6 GeV positrons.
In this analysis, a data sample of $23.6\,\mathrm{fb}^{-1}$ taken at 
the $\Upsilon$(5S) CM energy of $\sim$10869 MeV is used.
A data sample of $40.0\,\mathrm{fb}^{-1}$ taken in the continuum,
60 MeV in center-of-mass energy below the $\Upsilon$(4S),
and a data sample of $64.9\,\mathrm{fb}^{-1}$ taken at the
$\Upsilon$(4S) resonance energy, 
were also used for this analysis.
The continuum and $\Upsilon$(4S) data, which were 
collected with the same vertex detector,
provide comparable detector systematics.
All experimental conditions of data-taking at the $\Upsilon$(5S)
were nearly the same as at the $\Upsilon$(4S) or continuum runs
except for beam energies and luminosities.

The Belle detector is a general-purpose large-solid-angle magnetic
spectrometer that consists of a silicon vertex detector (SVD),
a central drift chamber (CDC), an array of
aerogel threshold Cherenkov counters (ACC), a barrel-like
arrangement of time-of-flight scintillation
counters (TOF), and an electromagnetic calorimeter
comprised of CsI(Tl) crystals (ECL) 
located inside a superconducting
solenoidal coil with a 1.5~T magnetic field.
An iron flux-return located outside the coil is
instrumented to detect $K^0_L$ mesons and to identify muons (KLM).
The detector is described in detail elsewhere~\cite{belle}.
A GEANT-based detailed simulation of the Belle detector is used
to produce Monte Carlo (MC) event samples and determine efficiencies.

\section{Selections}

Charged tracks are reconstructed using hit information from the CDC and SVD.
Kaon and pion mass hypotheses are assigned to the charged tracks using 
a likelihood ratio 
${\cal L}_{K/\pi} = {\cal L}_K/({\cal L}_K + {\cal L}_{\pi})$;
the likelihood ${\cal L}$ is obtained by combining 
information from the CDC (specific ionization ($dE/dx$)),
TOF, and ACC.
We require ${\cal L}_{K/\pi} > 0.6$ (${\cal L}_{K/\pi} < 0.6$) 
for kaon (pion) candidates~\cite{belle}.
The resulting identification efficiency varies 
from 86$\%$ to 91$\%$ (94$\%$ to 98$\%$) for kaons (pions),
depending on the momentum.

Standard procedures are used to identify leptons. Electrons
are identified using a combination of the specific ionization
measurement from the CDC, the ACC response, and the 
electromagnetic shower position, shape and energy measurements
from the ECL \cite{ele}.
The electron identification probability is required to exceed 0.5.
Muons are identified with KLM hit positions and penetration depth \cite{muo};
the standard muon prerejection cuts are applied and
the muon likelihood ratio is required to be larger than 0.8.
With these requirements, the combined reconstruction and identification
efficiency is flat in momentum and around $(75-78)\%$ for both 
electrons and muons with momentum larger than 1\,GeV/c.
The efficiency decreases with momentum in the lower momentum region.
Lepton tracks are required to satisfy $dr < 2\,$cm and 
$|dz| < 5\,$cm, where $dr$ and $|dz|$ are the
distances of closest approach to the nominal interaction point in the
plane perpendicular to the beam axis ($r-\phi$ plane) and along
the beam direction, respectively.
To remove leptons produced in $J/\psi$ decays and photon
conversions, we require
$|M(J/\psi) - M(e^+ h^-)|> 50\,$MeV/c$^2$,
$|M(J/\psi) - M(\mu^+ h^-)|> 50\,$MeV/c$^2$ and
$M(e^+ e^-) > 100\,$MeV/c$^2$, where $h^-$ is any
charged particle and $e^+$($\mu^+$) is any
identified electron (muon) reconstructed in the event.
Charge-conjugate combinations are also implied.
Finally, momentum requirements $P(e^+) > 0.5\,$GeV/c 
and $P(\mu^+) > 0.8\,$GeV/c in the center-of-mass (CM) system are applied,
and the polar angle requirements $17^\circ < \theta (e^+) < 150^\circ$ and 
$25^\circ < \theta (\mu^+) < 145^\circ$ in the laboratory system 
are applied to select regions of high reconstruction efficiency.  

Only the cleanest decay mode $D_s^+\,\to\,\phi \pi^+$ is used
to measure the $D_s^+$ meson yield.
The invariant mass of $\phi\,\to\,K^+ K^-$ candidates is
required to be within \mbox{$\pm$ 12\,MeV/$c^2$} of the nominal $\phi$ mass.
The $\phi$ helicity angle distribution is expected to be proportional 
to \mbox{cos$^2 \theta_{\rm hel}^{\phi}$}; therefore,
the requirement \mbox{$|$cos $\theta_{\rm hel}^{\phi}| > 0.25$} is applied.
The helicity angle $\theta_{\rm hel}^{\phi}$ is defined 
as the angle between the directions of the $K^+$ and $D_s^+$ momenta 
in the $\phi$ rest frame.
Although the $D_s^+$ mass resolution in the
kinematic region studied is only $(4-5)\,$MeV/c$^2$, we select
$\phi \pi^+$ combinations within $\pm 50\,$MeV/c$^2$
of the nominal $D_s^+$ mass, because sufficient sidebands are
needed to fit the $D_s^+$ signal.

The large background under the $D_s^+$ signal 
from $e^+e^- \rightarrow q \bar{q}$ continuum events 
($q = u,d,s,$ or $c$) is significantly
suppressed by the $D_s^+$ selection requirements and,
additionally, by the same-sign lepton requirement.
Therefore, to avoid uncertainties due to a variation of selection efficiency 
between the $\Upsilon$(5S), $\Upsilon$(4S) and continuum 
data samples, no topological cuts are applied in this analysis
to suppress continuum.

The $B_s^0$ mesons are produced at the $\Upsilon$(5S)
through the intermediate $B_s \bar{B}_s$, $B_s^* \bar{B}_s$, 
$B_s \bar{B}_s^*$ or $B_s^* \bar{B}_s^*$ channels,
with subsequent $B_s^* \to B_s \gamma$ decay.
These channels are not separated in this analysis; the
signal MC simulation generates 93$\%$ of events
in the $B_s^* \bar{B}_s^*$ channel,
and 7\% of events in the $B_s^* \bar{B}_s$ and $B_s \bar{B}_s^*$ channels. 
The $b\bar{b}$ production cross-section at the $\Upsilon$(5S)
has been measured in Ref. \cite{beli} to be 
$\sigma_{b\bar{b}}^{\Upsilon{\rm (5S)}} = 0.302 \pm 0.015$ nb, and
the fraction of $B_s^{(*)} \bar{B}_s^{(*)}$ events over all
$b\bar{b}$ events is taken from the PDG \cite{pdg}: $f_s = 
(19.5^{+3.0}_{-2.2})\%$.

\section{Number of \boldmath{$D_s^+$} mesons 
from \boldmath{$B_s^0$} decays}

To obtain the number of $D_s^+$ mesons produced from
$B_s^0$ decays, we compare the normalized $D_s^+$ momentum
distribution in the $\Upsilon$(5S), $\Upsilon$(4S)
and continuum data samples using a procedure
similar to that described in Ref.\cite{beli}.
The normalized momentum
is defined as $x(D_s^+) = P(D_s^+) / P_{\rm max}(D_s^+)$, where
$P(D_s^+)$ is the $D_s^+$ momentum in the CM,
and $P_{\rm max}(D_s^+)$ is the expected $D_s^+$ momentum
if the $D_s^+$ were produced in the process 
$e^+ e^- \to D_s^+ D_s^-$ at the same CM energy.

The comparison of the normalized momentum $x(D_s^+)$ distributions 
of the $\Upsilon$(5S) and continuum data samples (Fig.~1a)
and of the $\Upsilon$(4S) and continuum data samples (Fig.~1b)
shows excellent agreement in the region $x(D_s^+) > 0.5$, 
where $b\bar{b}$ events cannot contribute.
The continuum distributions are normalized using the 
energy-corrected luminosity ratio.
To obtain these distributions, the $D_s^+$ mass spectra are fitted
by a Gaussian to describe the signal and a linear function to describe
the background in each bin of $x(D_s^+)$.
The Gaussian width is fixed to the value obtained from
MC simulation; the Gaussian mean is fixed to that obtained from fitting
the $D_s^+$ signal in the whole $x(D_s^+) < 0.5$ range.
The signal normalization and the background parameters are allowed to float.

%=======================  \label{fig:figall}  =================================
\begin{figure}[htb]
\begin{center}
\includegraphics[width=7.cm,height=7.cm]{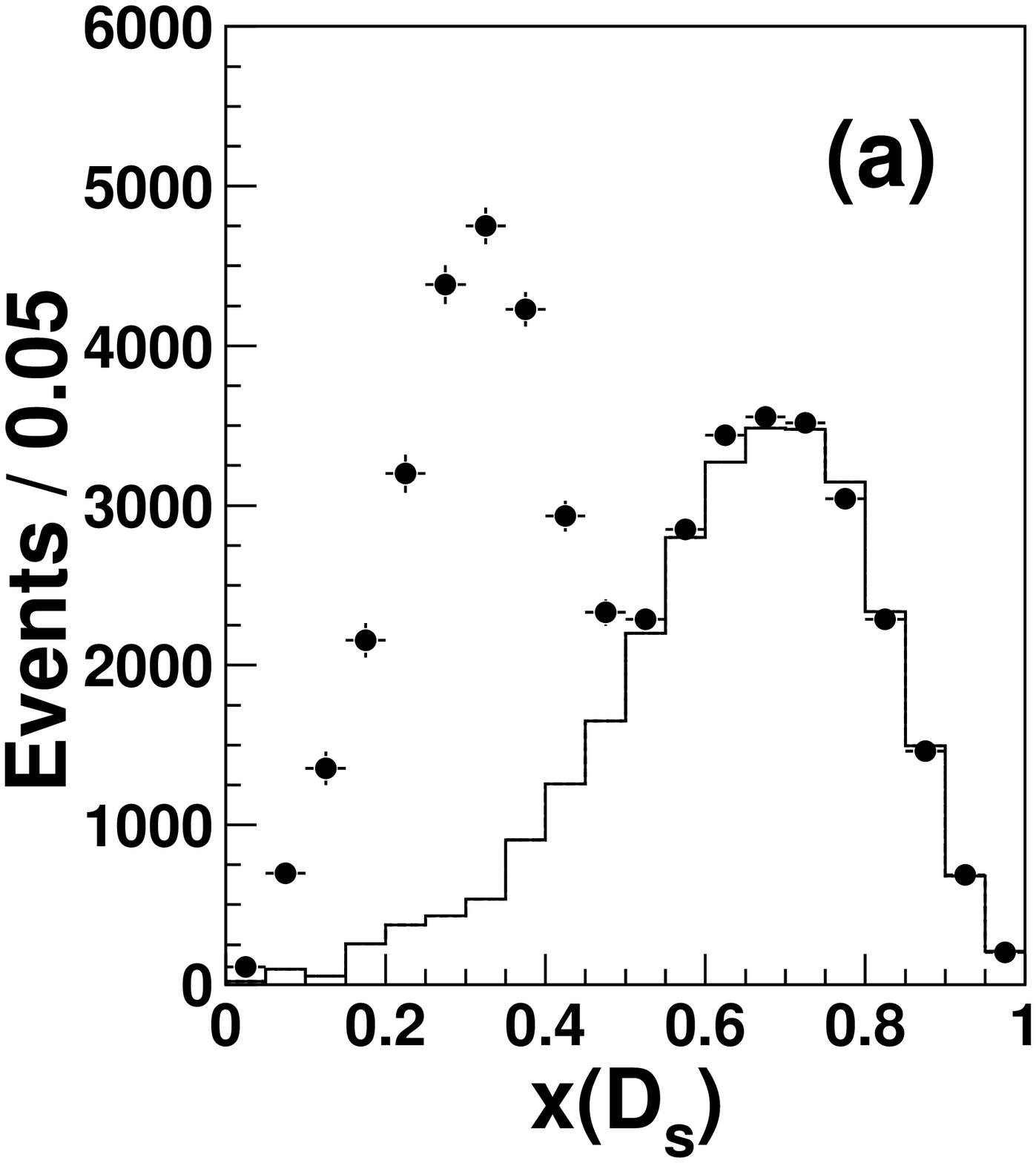}\includegraphics[width=7.cm,height=7.cm]{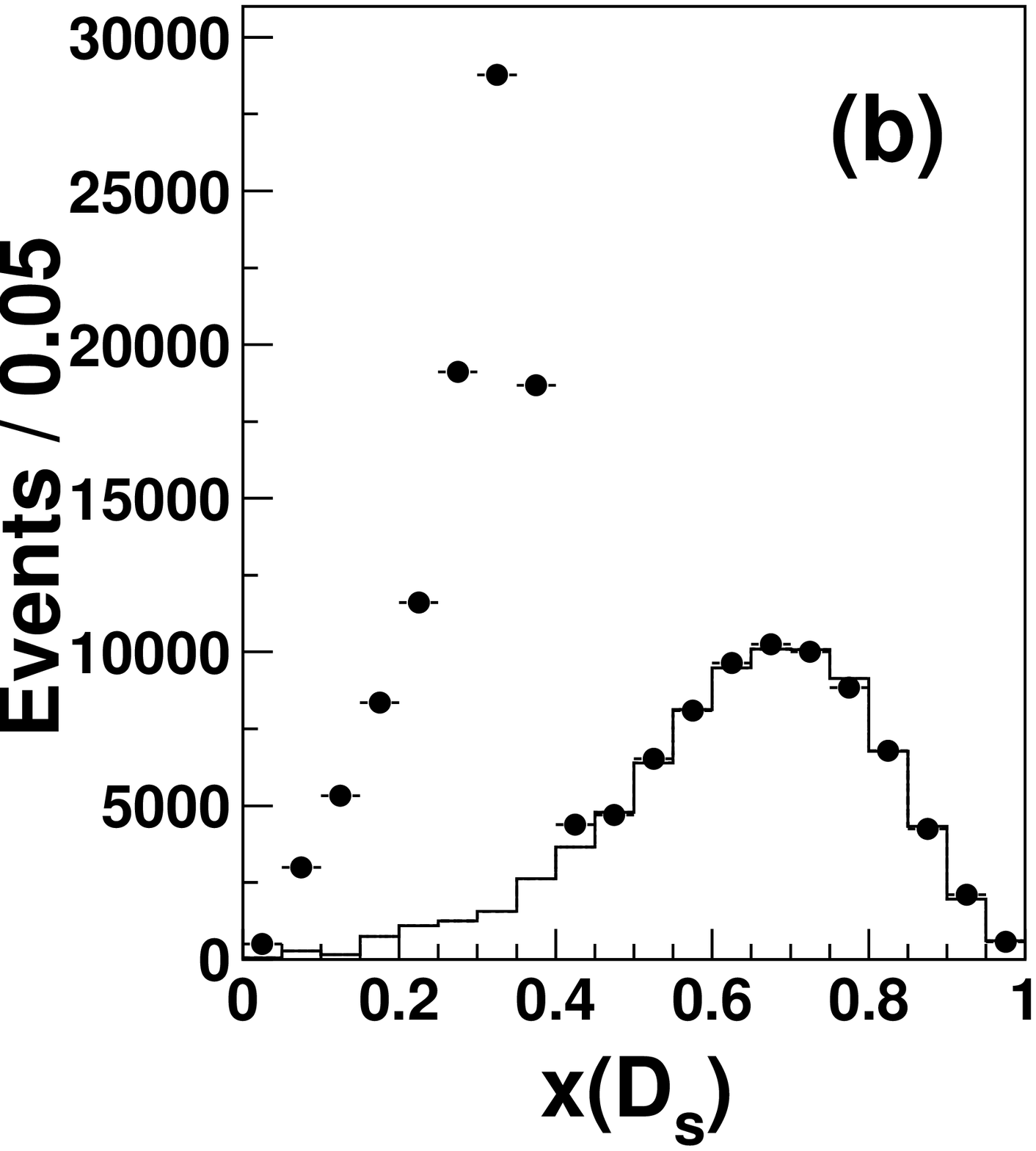}
\end{center}
\caption{Comparison of the $D_s^+$ normalized momenta $x(D_s^+)$ for the 
$\Upsilon$(5S) and continuum data samples (a), and for the
$\Upsilon$(4S) and continuum data samples (b).
The points with error bars are the
$\Upsilon$(5S) and $\Upsilon$(4S) data, while the histograms show the 
normalized continuum.}
\label{figa}
\end{figure}
%=======================  \end{fig:figall}  =================================

Continuum contributions were subtracted from the $x(D_s)$ distributions for 
the $\Upsilon$(5S) and $\Upsilon$(4S) data samples (Fig. 1),
and the obtained distributions
are used to calculate the number of $D_s^+$ mesons produced 
from $B_s^0$ decays.
The $\Upsilon$(4S) distribution is normalized to the number of
the $b\bar{b}$ pairs produced at the $\Upsilon$(5S) 
that do not hadronize to $B_s^0$ mesons.
To obtain this normalization, we use the number 
of $B\bar{B}$ pairs in our $\Upsilon$(4S) sample, 
$N^{\Upsilon{\rm (4S)}}(B\bar{B})=(70.95 \pm 0.79) \times 10^6$.
The number of $b\bar{b}$ pairs in the $\Upsilon$(5S) data sample
that do not hadronize to $B_s^0$ mesons is obtained from the formula:

\begin{equation}
N_{b\bar{b}}^{\Upsilon{\rm (5S)}}(B\bar{B})\ =\ {\cal L}_{\Upsilon{\rm (5S)}}\ \times \ \sigma_{b\bar{b}}^{\Upsilon{\rm (5S)}}\ \times \ ( 1- f_s )\
\end{equation} 

Finally, this normalization is calculated,
$N_{b\bar{b}}^{\Upsilon{\rm (5S)}}(B\bar{B}) / N^{\Upsilon{\rm (4S)}}(B\bar{B})\ = \ 1 / (12.37 \pm 0.79)$. 
We find that
about 2/3 of $D_s^+$ mesons produced at the $\Upsilon$(5S)
originate from $B_s^0$ decays. 
After the $B\bar{B}$ contribution
is subtracted and all bins are summed over the
interval $x(D_s) < 0.5$, the number of $D_s^+$ mesons
produced from $B_s^0$ decays is obtained:
$N(B_s^0 \to D_s^+ X) =$ \mbox{$13434 \pm 357 (stat) \pm 479 (syst)$}.

\section{Same-sign \boldmath{$D_s^+$} and lepton events 
in \boldmath{$\Upsilon$(5S)} sample}

The number of same-sign $D_s^+$ and lepton events
is obtained at the $\Upsilon$(5S), $\Upsilon$(4S)
and continuum data samples as a function of 
lepton momentum for the whole
$D_s^+$ momentum range $0\ <\ P(D_s^+) \ < 2.6\,$GeV/c.
To obtain these distributions, the $D_s^+$ mass spectra are fitted
in each bin of $P(\ell^-)$ in a manner similar to that described in the
previous section.

The following backgrounds must be subtracted from the $\Upsilon$(5S)
data to obtain the same-sign $D_s^+$ and lepton events from $B_s^0$ decays:
%a pure $B_s^0$ originated lepton momentum spectrum:
\begin{itemize}
\item[1] Background due to $\Upsilon$(5S)$\,\to B\bar{B}(X)$ 
decays. This is the dominant background, which is evaluated 
using the momentum-corrected $\Upsilon$(4S)$\,\to B\bar{B}$ data.
\item[2] Background due to continuum under the $\Upsilon$(5S) resonance.
This background is significantly suppressed by the lepton
requirement.
It is evaluated using the continuum data sample.
\item[3] Tracks misidentified as leptons in
$\Upsilon$(5S)$\,\to B_s^{(*)} \bar{B}_s^{(*)}$ events.
This background is important for low lepton momentum.
It is evaluated using MC simulation 
of $\Upsilon$(5S)$\,\to B_s^{(*)} \bar{B}_s^{(*)}$ events.
\item[4] Residual electrons from photon conversions, muons from 
charged kaon decays, and leptons from
$J/\psi$ decays in $\Upsilon$(5S)$\,\to B_s^{(*)} \bar{B}_s^{(*)}$ 
events.
This background is expected to be small and is also evaluated
using $\Upsilon$(5S)$\,\to B_s^{(*)} \bar{B}_s^{(*)}$ MC simulation.
\end{itemize}

The lepton momentum distribution obtained at the $\Upsilon$(4S)
was corrected to take into account the $B$ meson momentum difference
between the $\Upsilon$(5S) and $\Upsilon$(4S) samples; this resulted 
in a smearing (up to $\sim$300 MeV/c at high lepton momenta) with 
only a small global effect.
To subtract the continuum contribution, lepton momenta are scaled from
continuum to the $\Upsilon$(5S) energy using the same scale
factor as for the $x(D_s^+)$ distributions. 

The raw lepton momentum distributions for the $\Upsilon$(5S) sample 
are shown in Fig.\ 2, together with the backgrounds discussed above.
The background contributions are not large in the momentum region
greater than 1.2\ GeV/c and are subtracted from the 
raw $\Upsilon$(5S) lepton momentum distribution. A small correction
is also applied to the electron distribution to take into account
the radiative energy losses. Uncertainties in this correction 
are small and are included in the systematic error.
After the background subtraction and the radiative losses correction, the
final lepton momentum distributions are corrected for 
efficiencies. These efficiencies are obtained from MC simulation.

%=======================  \label{fig:figall}  =================================
\begin{figure}[htb]
\begin{center}
\includegraphics[width=7.cm,height=7.cm]{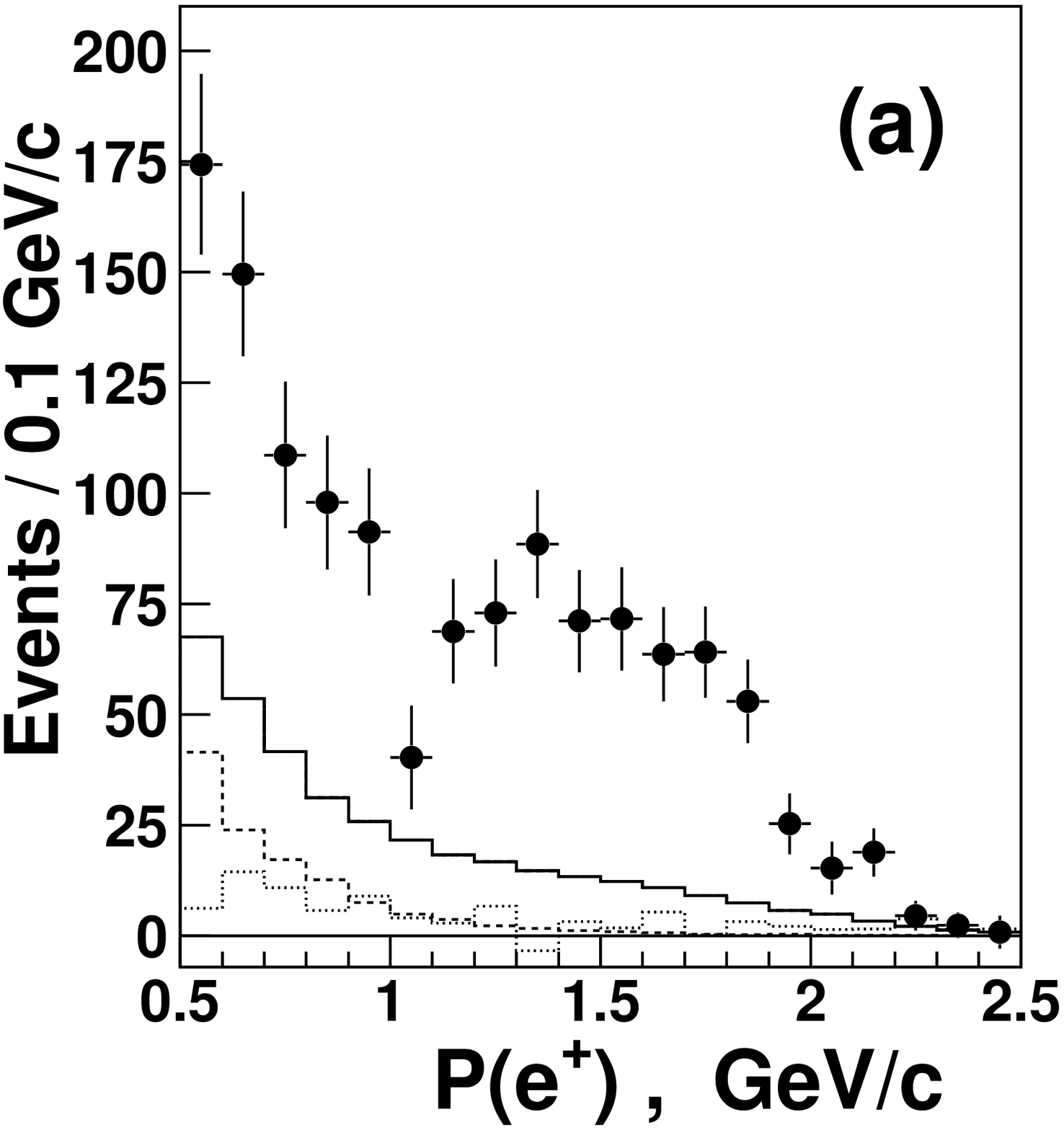}\includegraphics[width=7.cm,height=7.cm]{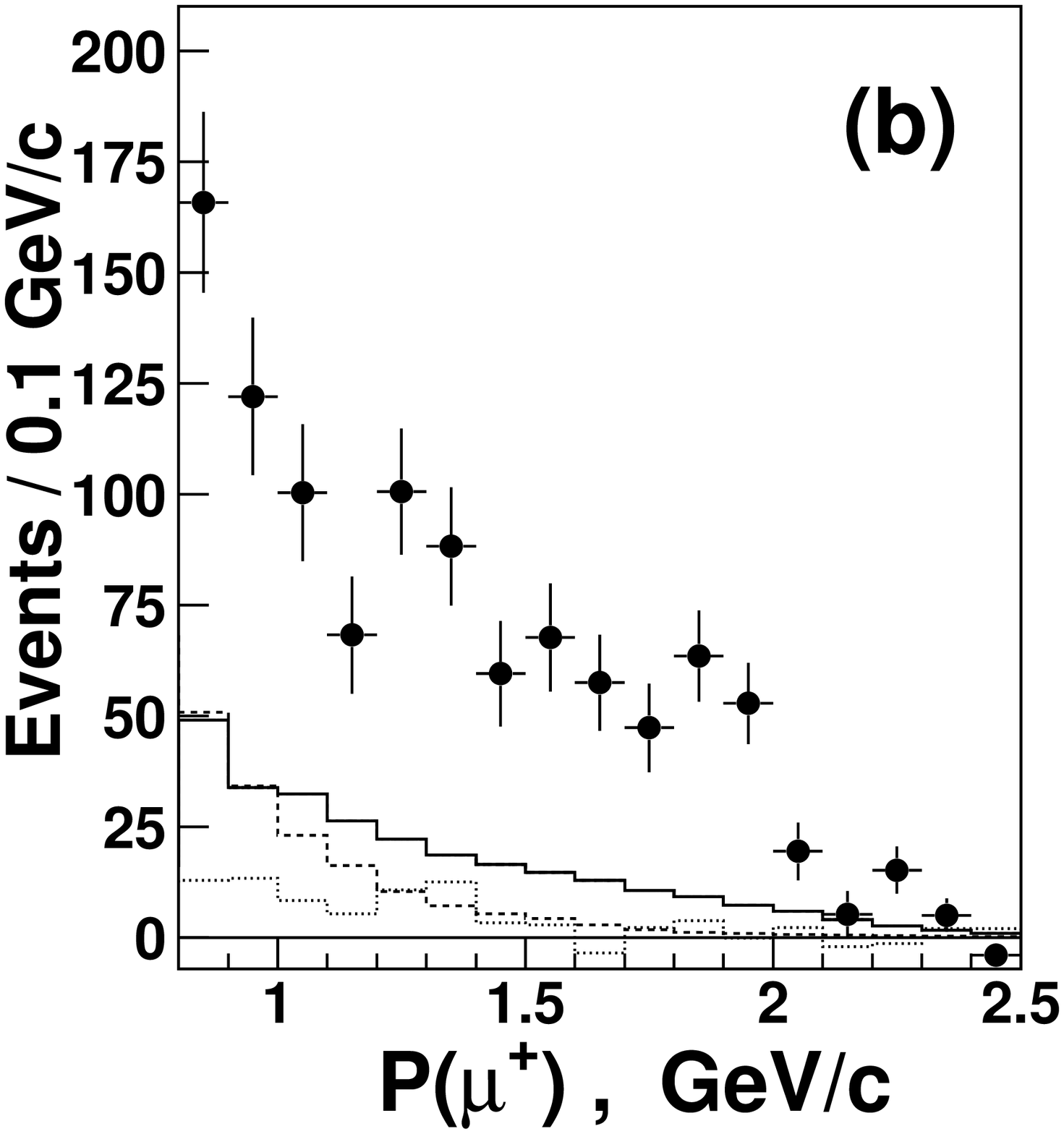}
\end{center}
\caption{Raw momentum distributions of electrons (a) and muons (b)
in the $\Upsilon$(5S) sample (data points with error bars),
together with estimates of the contributions from background (1)
(solid histograms), background (2) (dotted histograms) and backgrounds 
(3) and (4) (dashed histograms).}
\label{figb}
\end{figure}
%=======================  \end{fig:figall}  =================================

\section{Separation of primary and secondary leptons from \boldmath{$B_s^0$}
decays}

The final lepton momentum
distributions, after the background subtraction and efficiency
correction, are fitted with the sum of two contributions: one from
primary leptons and one from secondary leptons. Primary leptons,
produced directly in semileptonic $B_s^0$ decays, have a momentum
spectrum significantly harder than that of secondary leptons,
which are produced in the subsequent decays of $B_s^0$ daughters such as 
$D_s^+$, $D^0$ and $D^+$ mesons, and $\tau^+$ leptons. 
We expect a $(5-15)$\% contribution from
each of the latter three daughters relative to the $D_s^+$ contribution.
We take the $D_s^+$, $D^0$, $D^+$ and $\tau^+$ production ratios from 
our signal MC simulation to obtain the shape of the secondary lepton momentum
distribution, and we vary these ratios within reasonable limits. The
resulting systematic uncertainty is found to be small, because
the lepton spectra resulting from $D_s^+$, $D^0$, $D^+$ and $\tau^+$ decays 
are similar.

In the fit, we use fixed polynomial parametrizations for both the 
primary and the secondary lepton momentum spectra as determined from 
our MC simulation (Fig.\ 3),
which contains the most up-to-date knowledge of specific $B_s^0$ decays
based on experimental measurements and theoretical models.
Thus these shapes result from a semi-empirical description of $B_s^0$ decays. 

%=======================  \label{fig:figall}  =================================
\begin{figure}[htb]
\begin{center}
\includegraphics[width=7.5cm,height=7.cm]{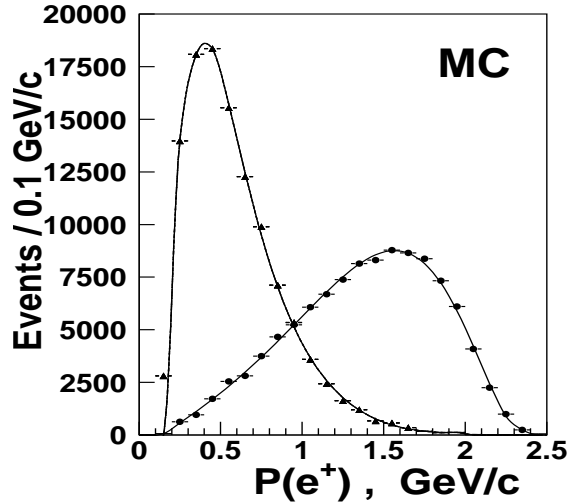}
\end{center}
\caption{Momentum distributions of primary (circles) and
secondary (triangles) electrons from $B_s^0$ decays in the MC simulation,
superimposed with the polynomial parametrizations used in the final
fit (curves).}
\label{fige}
\end{figure}
%=======================  \end{fig:figall}  =================================
%=======================  \label{fig:figall}  =================================
\begin{figure}[htb]
\begin{center}
\includegraphics[width=7.cm,height=7.cm]{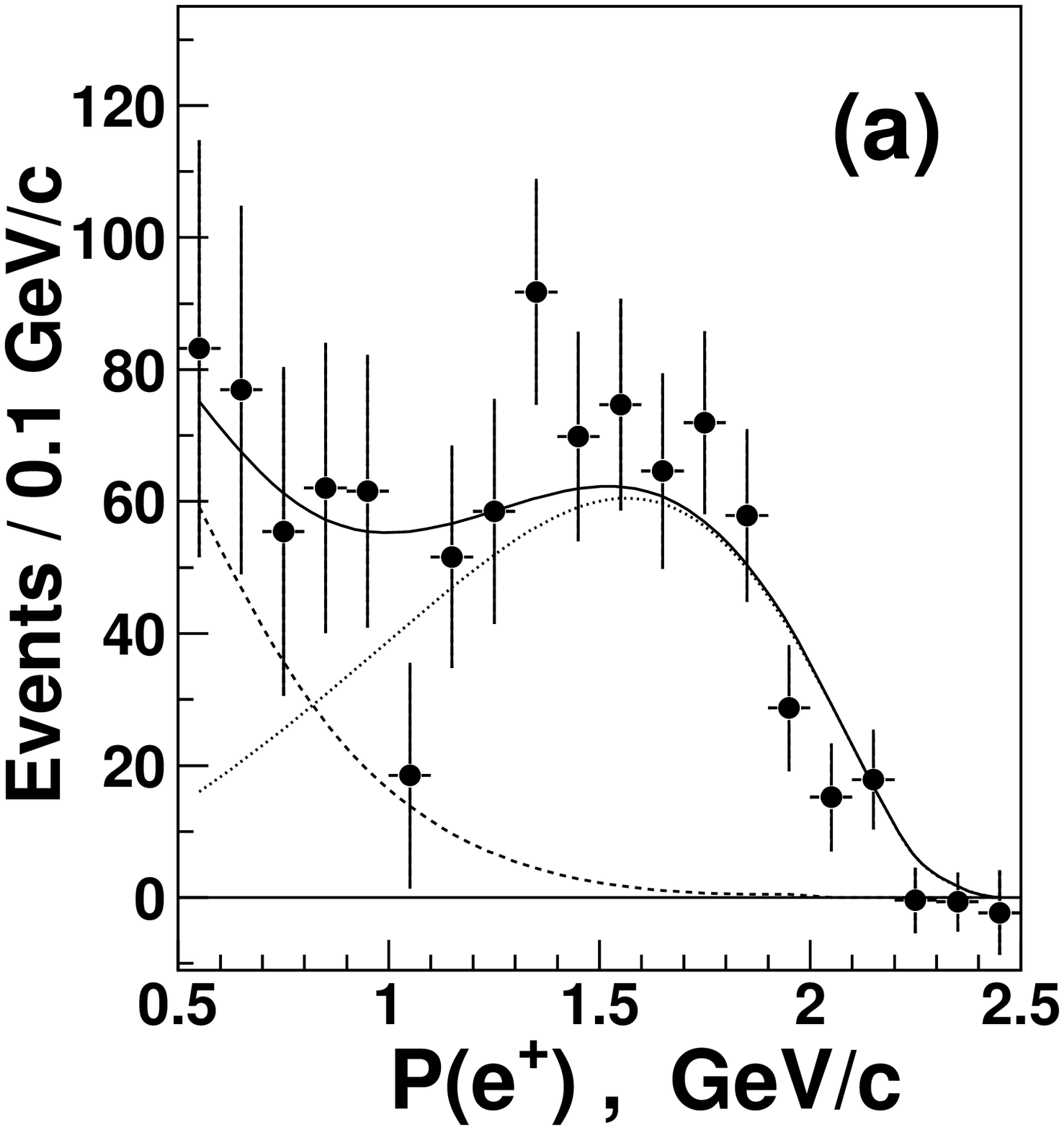}\includegraphics[width=7.cm,height=7.cm]{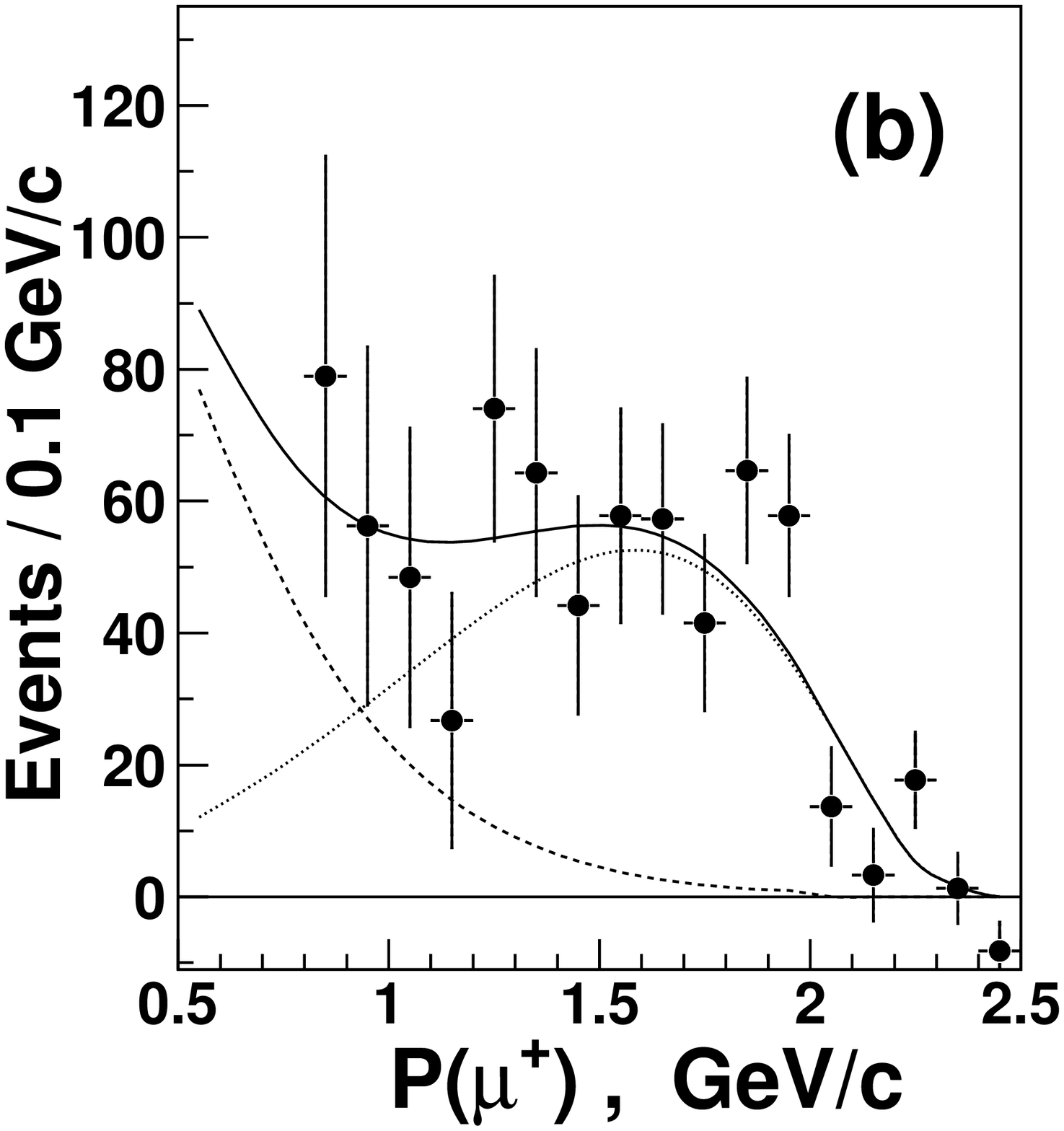}
\end{center}
\caption{The electron (a) and muon (b)
momentum distributions from $B_s^0$ decays. 
The solid curves show the results of the fits, and the dotted 
curves show the fitted contributions from primary and secondary leptons.}
\label{figc}
\end{figure}
%=======================  \end{fig:figall}  =================================

% using the $\Upsilon$(5S) data sample
The final leptonic momentum distributions produced 
in $B_s^0$ decays (Fig. 4)
obtained after background subtractions and an efficiency correction
(all corrections described above are also applied)
are used to extract the numbers of primary and secondary leptons.
We fit the data with a function which includes the sum of
two terms with fixed shapes and floating normalizations.
% The two shapes are fixed from MC simulation of
% the primary and secondary lepton momentum distributions, 
% as described above.
Based on the MC simulation, we expect that the
momentum ranges selected in our analysis will contain 96.7$\%$ of the
primary electrons ($P(e^+) > 0.5\,$GeV/c) and 89.4$\%$ of the primary
muons ($P(\mu^+) > 0.8\,$GeV/c).
The only free parameters in the fit are the
numbers of primary and secondary leptons. We fit separately the
electron spectrum and the muon spectrum, as shown in \mbox{Fig.\ 4}. We also
fit the two distributions simultaneously, assuming equal electron
and muon production rates in $B_s^0$ decays, for both primary and secondary
leptons. The results of the three binned $\chi^2$-fits are given in
Table 1. They represent our measured numbers of a same-sign leptons 
accompanying $D_s$ meson in events with $B_s^0$ pairs 
from the $\Upsilon$(5S) sample.
% Although the numbers of secondary leptons have large errors,
% it does not strongly affect the numbers of primary leptons
% because of the different momentum distribution shapes.
% The fit yields $N(l) = 683 \pm 46$ events. These numbers correspond
% to the numbers of the same-sign $D_s^+$ and lepton events
% produced from $B_s^0$ pairs in the $\Upsilon$(5S) data sample.

\renewcommand{\arraystretch}{1.2}
\begin{table}[htb]
\caption{Results of the electron, muon and combined momentum spectra fits.
Numbers of leptons extracted from the fits correspond to the whole momentum 
region.}
\begin{center}
\vspace{0.2cm}
\label{tab:bfr2}
\begin{tabular}
{@{\hspace{0.3cm}}l@{\hspace{0.3cm}} @{\hspace{0.3cm}}c@{\hspace{0.3cm}} @{\hspace{0.3cm}}c@{\hspace{0.3cm}} @{\hspace{0.3cm}}c@{\hspace{0.3cm}}}
% @{\hspace{0.1cm}}c@{\hspace{0.1cm}} @{\hspace{0.3cm}}c@{\hspace{0.1cm}} @{\hspace{0.1cm}}c@{\hspace{0.1cm}} @{\hspace{0.3cm}}c@{\hspace{0.1cm}} }
% ----------------------------
\hline \hline
 Subsample & Number of & Number of & Fit, \\
  & primary leptons & secondary leptons & $\chi^2 / n. d. f.$ \\
\hline
Electron &  $731 \pm 63$ & $430 \pm 132$ & 0.99 \\
Muon &  $619 \pm 75$ & $583 \pm 334$ & 1.42 \\
Combined &  $683 \pm 46$ & $460 \pm 123$ & 1.16 \\
\hline \hline
\end{tabular}
\end{center}
\vspace{-0.1cm}
\end{table}

Finally, dividing these numbers by the full number of
$D_s^+$ mesons from $B_s^0$ decays in our $\Upsilon$(5S)
data sample (see Section 3) and applying an additional 
factor of 2 due to the $B_s^0$ mixing effect,
we obtain the following semileptonic branching fractions:

\begin{equation}
{\cal B}(B_s^0 \rightarrow X^+ e^- \nu)\, = (10.9 \pm 1.0 \pm 0.9)\%
\end{equation}
\begin{equation}
{\cal B}(B_s^0 \rightarrow X^+ \mu^- \nu)\, = (9.2 \pm 1.0 \pm 0.8)\%
\end{equation}
\begin{equation}
{\cal B}(B_s^0 \rightarrow X^+ \ell^- \nu)\, = (10.2 \pm 0.8 \pm 0.9)\% ,
\end{equation}
\noindent
where the last one represents an average over electrons and muons.

Table 2 lists the systematic uncertainties, 
which are combined in quadrature
to obtain the total systematic uncertainty. 
% The systematic uncertainty on the full $D_s^+$ meson number
% is dominated by the
% $\sigma_{b\bar{b}}^{\Upsilon{\rm (5S)}}$ and $f_s$ uncertainties;
% details are described in the text. 
Uncertainties due to continuum and $B\bar{B}$ background subtraction
are estimated by varying the measured normalization factors
by $1.5\,\sigma$; the shapes of the distributions are not varied.
Uncertainties due to the shapes of the primary and secondary lepton 
momentum spectra are evaluated by varying them within the expected 
uncertainties in the $B_s^0$ decay branching fractions.

The obtained branching fractions can be compared with the PDG value
${\cal B}(B^0 \rightarrow X^+ \ell^- \nu)\, = (10.33 \pm 0.28)\%$,
which is theoretically expected to be approximately the same,
neglecting a small possible lifetime difference and
small corrections due to electromagnetic and
light quark mass difference effects.

\renewcommand{\arraystretch}{1.2}
\begin{table}[htb]
\caption{List of systematic uncertainties on
the $B_s^0 \rightarrow X^+ \ell^- \nu$ branching
fraction measurements.}
\vspace{0.2cm}
\label{tab:bfr2}
\begin{tabular}
{@{\hspace{0.1cm}}l@{\hspace{0.1cm}} @{\hspace{0.1cm}}c@{\hspace{0.1cm}} }
% @{\hspace{0.1cm}}c@{\hspace{0.1cm}} @{\hspace{0.3cm}}c@{\hspace{0.1cm}} @{\hspace{0.1cm}}c@{\hspace{0.1cm}} @{\hspace{0.3cm}}c@{\hspace{0.1cm}} }
% ----------------------------
\hline \hline
 Source & Relative uncertainty \\
\hline
1. Lepton misidentification & $1\%$ \\
2. Signal and background shapes in $D_s^+$ mass fit & $3\%$ \\
3. Determination of full $D_s^+$ meson number & $4\%$ \\
4. Continuum background modeling & $2\%$ \\
5. $B\bar{B}$ background modeling & $2\%$ \\
6. MC lepton efficiency determination & $4\%$ \\
7. Variation of $D^0$, $D^+$ and $\tau^+$ contributions& $1\%$ \\ 
8. Shapes of primary and secondary lepton momentum spectra & $5\%$ \\ 
\hline
\ \ \ \ Sum in quadrature & $8.5\%$ \\
\hline \hline
\end{tabular}
\vspace{-0.1cm}
\end{table}

\section{Conclusions}

The total inclusive semileptonic $B_s^0 \rightarrow X^+ \ell^- \nu$
decay branching fraction is measured for the first time
using a 23.6\,fb$^{-1}$ data sample collected
at the $\Upsilon$(5S) resonance with the Belle
detector at the KEKB asymmetric energy $e^+ e^-$ collider.
The obtained value,
${\cal B}(B_s^0 \rightarrow X^+ \ell^- \nu)\, = (10.2 \pm 0.8 \pm 0.9)\%$,
is in a good agreement with the total inclusive semileptonic $B^0$
decay branching fraction.

\section{Acknowledgments}

We thank the KEKB group for the excellent operation of the
accelerator, the KEK cryogenics group for the efficient
operation of the solenoid, and the KEK computer group and
the National Institute of Informatics for valuable computing
and Super-SINET network support. We acknowledge support from
the Ministry of Education, Culture, Sports, Science, and
Technology of Japan and the Japan Society for the Promotion
of Science; the Australian Research Council and the
Australian Department of Education, Science and Training;
the National Science Foundation of China and the Knowledge
Innovation Program of the Chinese Academy of Sciences under
contract No.~10575109 and IHEP-U-503; the Department of
Science and Technology of India; 
the BK21 program of the Ministry of Education of Korea, 
the CHEP SRC program and Basic Research program 
(grant No.~R01-2005-000-10089-0) of the Korea Science and
Engineering Foundation, and the Pure Basic Research Group 
program of the Korea Research Foundation; 
the Polish State Committee for Scientific Research; 
%-> remove for now: under contract No.~2P03B 01324; 
the Ministry of Education and Science of the Russian
Federation and the Russian Federal Agency for Atomic Energy;
the Slovenian Research Agency;  the Swiss
National Science Foundation; the National Science Council
and the Ministry of Education of Taiwan; and the U.S.\
Department of Energy.

\end{document}